\documentclass[11pt,preprint]{aastex}

\usepackage{graphicx}

\usepackage{natbib}

\slugcomment{Revised 2013 April 1}

\shorttitle{Phaethon Outburst}
\shortauthors{Li and Jewitt}

\begin{document}


\title{Recurrent Perihelion Activity in (3200) Phaethon }

\author{Jing Li\altaffilmark{1} and David Jewitt\altaffilmark{1,2} }
\affil{(1) Department of Earth and Space Sciences,  University of California at Los Angeles}
\affil{(2) Department of Physics and Astronomy, University of California at Los Angeles}

\email{jli@igpp.ucla.edu}

\begin{abstract}
We present a study of planet-crossing asteroid (3200) Phaethon at three successive perihelia in 2009, 2010 and 2012, using the NASA STEREO spacecraft.  Phaethon is clearly detected in  2009 and 2012, but not in 2010.  In both former years, Phaethon brightened unexpectedly by $\sim$1 magnitude at large phase angles, inconsistent with the $\sim$1 magnitude of steady fading expected from a discrete, macroscopic body over the same phase angle range.   With a perihelion distance of 0.14 AU and surface temperatures up to $\sim$1000 K, a thermal origin of this anomalous brightening is strongly suspected.  However, simple thermal emission from Phaethon is too weak, by a factor $>$10$^3$, to explain the brightening.  Neither can ice survive on this body, ruling out comet-like sublimation.  Our preferred explanation is that brightening occurs as a result of dust produced and ejected from Phaethon, perhaps by thermal fracture and/or thermal decomposition of surface minerals when near perihelion.  A contribution from prompt emission by oxygen released by desiccation of surface minerals cannot be excluded.  We infer an ejected mass of order 
4$\times$10$^8 a_{mm}$ kg per outburst, where $a_{mm}$ is the mean dust radius in millimeters.  For plausible dust radii, this mass is small compared to the estimated mass of Phaethon ($\sim$2$\times$10$^{14}$ kg) and to the mass of the Geminid stream (10$^{12}$ kg to 10$^{13}$ kg) with which Phaethon is dynamically associated.  Perihelion mass-loss events like those observed in 2009 and 2012 contribute to, but do not necessarily account for the Geminids stream mass. 
\end{abstract}

\keywords{asteroids, comets}

\section{Introduction}
Object (3200) Phaethon (formerly 1983 TB) is dynamically associated with the Geminid meteor stream, suggesting that it is the long-sought parent of this stream \citep{1983IAUC.3881....1W}.  Unlike the cometary parents of most streams \citep[e.g.][]{2008EM&P..102..505J}, however, Phaethon has a distinctly asteroidal orbit (semimajor axis = 1.271 AU, eccentricity = 0.890, inclination = 22.2\degr), with an asteroid-like Tisserand parameter with respect to Jupiter \citep{1980M&P....22...83K} of $T_J$ = 4.5.  Phaethon is about 5 km in diameter \citep[V-band geometric albedo = 0.17,][]{1984LPI....15..878V} and appears to be dynamically associated not just with the Geminids, but also with at least two smaller (kilometer-scale) asteroids, namely 2005 UD \citep{2006A&A...450L..25O, 2006AJ....132.1624J, 2007A&A...466.1153K} and 1999 YC \citep{2008AJ....136..881K,2008M&PSA..43.5055O}. Together, these objects constitute the so-called ``Phaethon-Geminid Complex'' (PGC) presumably formed by the disruption of a larger parent body.  Physical observations show that the PGC members share similar neutral-blue colors that are relatively rare in the main-belt population and suggesting a common composition. The PGC objects all possess small perihelion distances (0.14 AU in the case of Phaethon) resulting in high surface temperatures.

Physical observations of Phaethon when far from perihelion have consistently failed to show evidence for on-going mass loss, either in gas or in scattered continuum from entrained dust \citep{1984Icar...59..296C,1996Icar..119..173C,2005ApJ...624.1093H,2008Icar..194..843W}.  Neither do the associated bodies 2005 UD and 1999 YC (references above) show evidence for on-going mass loss.  However, observations of Phaethon at very small solar elongations using the STEREO spacecraft revealed anomalous photometric behavior near perihelion in 2009 \cite[][hereafter, Paper I]{2010AJ....140.1519J}.  Specifically, Phaethon was observed to brighten with increasing phase angle (near $\sim$80\degr) when near perihelion, a behavior inconsistent with scattering from any macroscopic solid body and opposite to the phase functions of known solar system objects \citep{1973AJ.....78..267L,2007Icar..187...41L,2007Icar..188..195L,2009Icar..204..209L}.  We interpreted the brightening as caused by an increase in the scattering cross-section following the ejection of dust from Phaethon.  The required mass of dust is $\sim$2.5$\times$10$^{8} a_{mm}$ kg, where $a_{mm}$ is the unmeasured size of the dust grains, expressed in millimeters (Paper I).  These observations are potentially important both for showing that Phaethon is an active source of matter for the Geminid stream, and for illuminating physical processes induced on bodies when close to the Sun. 

In this paper, we report new STEREO observations of Phaethon in 2010 and 2012 combined with a re-analysis of measurements from 2009 earlier reported in Paper I.  The principal question we seek to address is whether the anomalous brightening detected in 2009 is recurrent at subsequent perihelia.

\section{Observations}
The present observations were made with the Heliospheric Imagers (HI) which are part of the Sun Earth Connection Coronal and Heliospheric Investigation (SECCHI) package \citep{2008SSRv..136...67H,2009SoPh..254..387E} onboard the STEREO spacecraft.  The HI instruments consist of two wide-angle visible-light imagers, HI-1 and HI-2, with field centers offset from the solar center by 14.0$^\circ$ and 53.7$^\circ$, and fields of view $20^\circ$ and $70^\circ$, respectively.  The HI detectors are charge-coupled devices (CCDs) with 2048$\times$2048 pixels. These  are usually binned onboard to 1024$\times$1024 pixels, resulting in a binned pixel angular size of 70\arcsec~for HI-1 and 4\arcmin~for HI-2.  The very large pixels subtend solid angles 10$^5$ to 10$^6$ times those of pixels commonly used on night-time telescopes, but this is an advantage for the intended detection of large scale, diffuse structures in the solar corona.

We used the standard HI-1 camera Level 1 images in our study. The data are publicly available via the UK Solar System Data Center (UKSSDC) web site. \notetoeditor{http://www.ukssdc.rl.ac.uk/solar/stereo/data.html} Multiple short-exposure images are taken before a 1024$\times$1024 pixel-image is transmitted to Earth. The individual exposure time is 40 seconds, the exposure cadence is 60 seconds and 30 images are combined to make a single image having an exposure time of 20 minutes. One such image is obtained every 40 minutes.  This strategy is designed to remove cosmic rays, achieve statistical accuracy and avoid saturation of the background corona.  The exposure sequence is chosen so that the drift of the star field through the field of view (at $\sim 150$\arcsec~hr$^{-1}$) is comparable to the binned pixel size. The quantum efficiency of the CCD camera and the absolute transmission efficiencies of the optics are nearly constant across the  6300 to 7300 \AA~wavelength range (Eyles et al.~2009).

\subsection{2009 and 2012 Observations}
The short orbit period (1.43 yr) offers frequent opportunities to observe Phaethon at perihelion.  The orbit period of STEREO A is $\sim$346 days so that three STEREO orbits (2.84 yr) are almost exactly equal to twice the orbit period of Phaethon (2.86 yr). Therefore, observations in 2009 (perihelion June 20) and 2012 (perihelion May 02) share similar perspectives viewed from STEREO A.  

\begin{figure}
\epsscale{1.0}
\begin{center}
\includegraphics[width=1.0\textwidth]{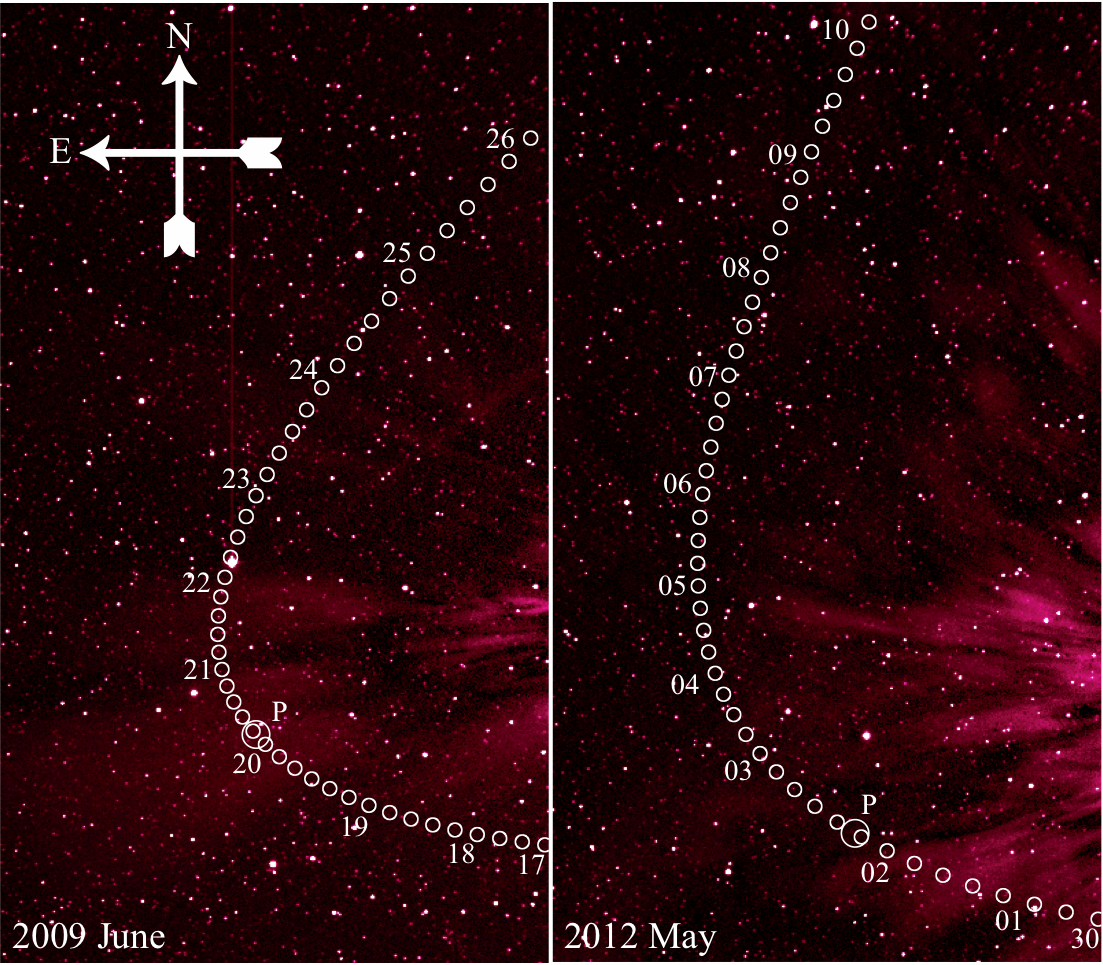}
\caption{The path of Phaethon (white circles) across the field of view of the STEREO A HI-1 camera in 2009 (left) and 2012 (right). Numbers along the path show the day of the month. The perihelia are indicated with the letter ``P'' and a large circle. The sun is on the right. The images were taken on the dates when Phaethon was at the perihelion. Both panels show  400$\times$700 pixels (7.8\degr$\times$13.6\degr). \label{path} }
\end{center} 
\end{figure}

We obtained the celestial coordinates of Phaethon from NASA's HORIZONS Web-Interface and transformed them to pixel coordinates on the HI cameras in the ``AZP''  Zenithal projection \citep[][Thompson, private communication 2012]{2002A&A...395.1077C, 2006A&A...449..791T}.   The sky-plane trajectories of Phaethon are shown in Fig. (\ref{path}). As expected, they follow similar paths  in 2009 and 2012. The angular speed of Phaethon in the images is the result of the combined spacecraft parallactic motion and object Keplerian motion. Viewed from STEREO A, Phaethon moved  relative to the field center at a speed varying between 0\arcsec~hr$^{-1}$ and 175\arcsec~hr$^{-1}$ from east to west, and at a roughly constant speed 180\arcsec~hr$^{-1}$ (in 2009) and 200\arcsec~hr$^{-1}$ (in 2012) from south to north.  Phaethon was readily apparent even in a cursory visual examination of the 2009 and 2012 data.

\subsection{2010 Observations}
\begin{figure}
\epsscale{0.9}
\begin{center}
\includegraphics[width=0.6\textwidth]{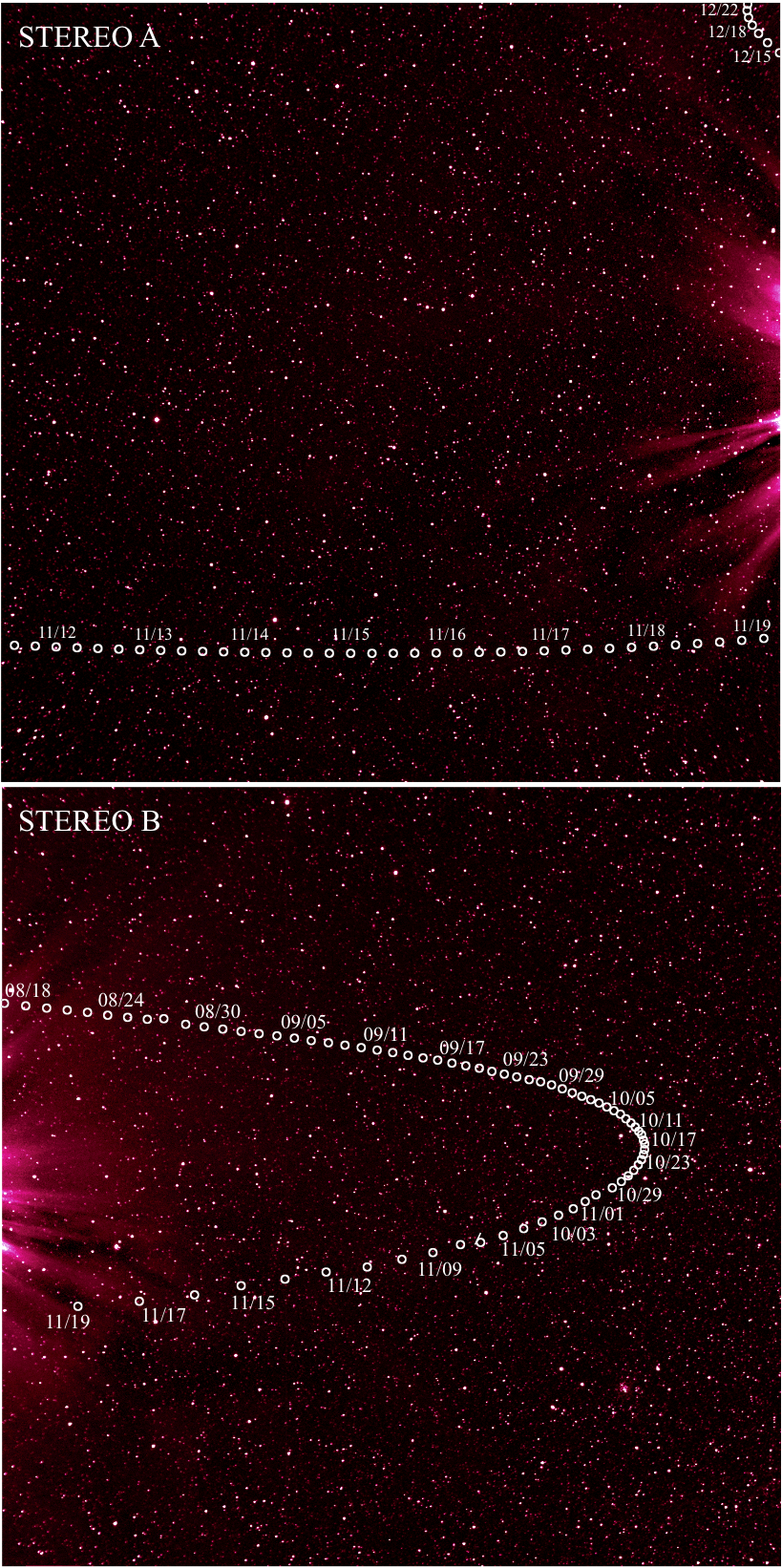}
\caption{The path of Phaethon (white circles) across the STEREO A (upper) and B (lower) HI-1 fields of view in 2010. Numbers along the path show the date in month/day format. Perihelion occurred on UT 2010 November 25 18:00, at which time Phaethon was not within the field of either STEREO camera. Panels show the full size HI-1 images of 20\degr$\times$20\degr. The sun is on the right in the upper panel, and on the left in the lower panel. \label{path2010} }
\end{center} 
\end{figure}

The observing geometry in the intervening orbit in 2010 is completely different from that in 2009 and 2012. Phaethon was in the field of view of STEREO A from November 11-19 and from December 15-25  (upper panel in Figure \ref{path2010}). It left the HI-1 field of view six days before its perihelion on November 25 and only re-entered the field about twenty days after perihelion in the second time period. Phaethon was not detected in either observing window. During the first time interval, the phase angle was above $120^\circ$ making an unfavorable  condition for detection. During the second time interval, the phase angle was a modest 10$^\circ$, but the heliocentric distance had increased to $\sim$0.8 AU. As a result, the nominal predicted magnitude had fallen to V $\sim$15, making Phaethon too faint to be detected in STEREO data. 

Phaethon stayed within the field of the STEREO B HI-1 camera for almost three months during its 2010 return (bottom panel in Fig. \ref{path2010}) but still contrived to leave the field of view six days before perihelion. Between mid-August and mid-November, the phase angle varied from $\sim0^\circ$ to a maximum $30^\circ$, while the heliocentric distance decreased from 1.8 to 0.3 AU. The apparent visual magnitude calculated assuming solid body reflection decreased from 18.5 to 12.5.  At its predicted brightest, Phaethon should have been marginally within reach of STEREO HI-1 camera detection, but in practice the object was not detected.

Presumably as a result of these different observational circumstances, the anomalous brightening detected in 2009 and 2012 was not found in 2010. Activity specifically at perihelion could not be detected because Phaethon was outside the fields of both the A and B cameras when at perihelion.   We conclude that differences in the observational geometry prohibited the detection of Phaethon and its activity at perihelion in 2010. 

\subsection{Keck Observations}
To supplement the near-Sun photometry from STEREO, we also observed Phaethon when far from the Sun using the Keck 10-m telescope on UT 2012 October 14.5.  We used the Low Resolution Imaging Spectrometer (LRIS, see Oke et al. 1995) and a broadband R filter (central wavelength $\lambda_c$ = 6417\AA, full-width at half maximum, FWHM  = 1185\AA) in seeing of FWHM = 1.0\arcsec.  The LRIS pixel scale is 0.135\arcsec pixel$^{-1}$.   Flat field images were obtained using an illuminated patch on the inside of the Keck dome. The data were photometrically calibrated using solar-colored standard stars from \citet{1992AJ....104..340L}.  We measured $R$ = 17.25$\pm$0.05 at  heliocentric and geocentric distances 2.184 AU and 1.346 AU, respectively, and phase angle $\alpha$ = 18.1\degr.  This measurement refers to an unknown rotational phase of Phaethon. 

\section{Photometry}
With pixel sizes 70\arcsec, images of HI-1 are sensitive to the diffuse background coronal emission but  under-sample the point-spread function of the telescope, and lead to frequent confusion with background sources as Phaethon moves across the sky.  In these data, the relatively uniformly distributed background corona overwhelms both field stars and Phaethon. To suppress the coronal background, we grouped images over timespans from a few hours to a few days. Within each group, the background corona was relatively constant, and was calculated using the minimum filter technique (an IDL code ``min\_filter'' in the SolarSoft IDL package was used) and then subtracted from the images in the group. The coronal filtering leaves rapidly varying background structures visible in the images, but they are muted in intensity compared to those present in the raw data.  Sample corona-subtracted images are shown in Figures (\ref{path}) and (\ref{path2010}).  

During the 30 minutes required to accumulate a single HI-1 image, Phaethon moved a maximum of 1.8 pixels (126\arcsec) relative to the CCD. We experimented with different photometry apertures in order to examine the effects of trailing and background subtraction.  In particular, we checked that the measured brightness variations are not related to the angular speed of Phaethon and therefore to trailing of the images.  Very small apertures are sensitive to the trailing, while very large ones achieve poor signal-to-noise ratios owing to uncertainty in the large background signal from the corona.  By trial and error, we chose a photometry box of 5$\times$5 pixels (350\arcsec$\times$350\arcsec). The sky background was obtained from the median count in 56 sky pixels defined by a box 9$\times$9 pixels (630\arcsec $\times$630\arcsec) wide, surrounding the 5$\times$5 object extraction box.  Results obtained using larger photometry apertures were consistent with those eventually used, but with a larger uncertainty due to noise in the coronal background.   To facilitate comparison with measurements from Paper I and to make sure that no systematic effects were introduced during our analysis, we elected to completely re-reduce the 2009 data-set, as well as those obtained near perihelion in 2010 and 2012.  

\begin{figure}
\epsscale{1.0}
\begin{center}
\includegraphics[width=0.6\textwidth]{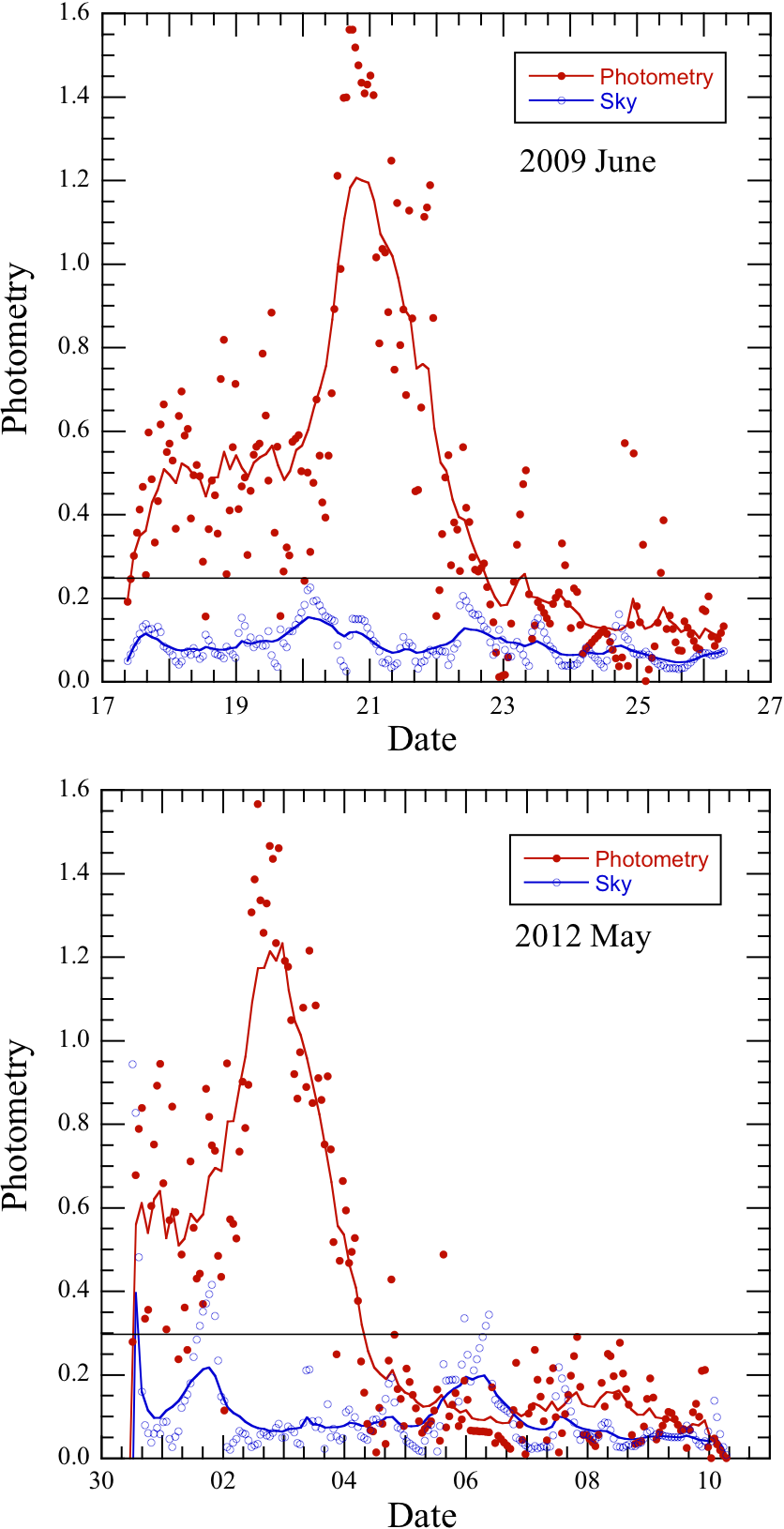}
\caption{Raw photometry of Phaethon (red circles) as a function of time near perihelion in 2009 and 2012. The median sky brightness surrounding Phaethon is shown with blue circles. Smoothed curves (red and blue lines) have been plotted to guide the eye. Horizontal lines represent the photometry thresholds that are three times of the sky mean (see Table \ref{stats}). Above the levels, Phaethon was detected. \label{photo} }
\end{center} 
\end{figure}

The extracted, raw Phaethon photometry from 2009 and 2012 is shown in Figure (\ref{photo}).  The photometry statistics are given in Table (\ref{stats}).  To be conservative, we set a photometry threshold at three times the mean of the sky, as shown by the horizontal lines in Figure (\ref{photo}). The thresholds lead to windows in which the photometry is useful lasting for 5.5 days (from June 17.5 to 23) in 2009 and 3.8 days (from April 30.5 to May 4.3) in 2012. The corresponding phase angles range from $29^\circ$ to $130^\circ$ in 2009, and from $32^\circ$ to $105^\circ$ in 2012.  In both years, Phaethon shows an apparent brightness surge by a factor of $\sim$3 relative to the pre-surge brightness and lasting for approximately two days.  Each surge is characterized by a sudden rise and fall, with no ``plateau'' phase in between. Phaethon's brightness surges are too large and too long-lived (the photometry aperture crossing time is measured in hours while the brightennings last for days) to be caused by passing stars.  Contaminating stars of sufficient brightness would, in any case, be apparent in visual examination of the images but were not seen. In fact, we discarded images in which bright stars interfered with the object. Furthermore, the background is constant with respect to time except for small excursions of $\pm$10\% due to passing stars (blue circles in Figure \ref{photo}).   For these reasons, we are confident that the brightening in 2009 and 2012 is real and associated with Phaethon, not caused by background object contamination or sky subtraction errors.   
 
To photometrically calibrate the measurements of Figure (\ref{photo}), we chose field stars near the projected path of Phaethon across the CCD. Field stars were chosen to be within 20 pixels ($\sim 23.3$\arcmin) of Phaethon,  with magnitudes $8 \le V \le$ 10 and of spectral types FGK. Eleven standard stars in 2009, and fifteen stars in 2012 were available for the Phaethon flux calibration (see Table \ref{stars}). Reference star photometry was obtained in the same manner as for Phaethon.   The 6300 \AA~to 7300 \AA~passband of the STEREO camera is close to astronomical R-band.  However, Phaethon is nearly neutral and the reference stars were selected to be similar in color to the Sun.  Therefore, our measurements of the brightness ratio are equivalent to V-band measurements save for a small, color-correction offset $\lesssim$0.1 mag., which is negligible compared to the uncertainties of measurement.

Figure (\ref{magv}) shows the resulting  apparent magnitudes of Phaethon as a function of time in the time windows of the Phaethon visibility. The times of perihelion UT 2009 June 20 07:12 and UT 2012 May 2 07:12 are indicated by arrows and marked with the letter ``P''.  Shown for comparison are the magnitudes predicted by the JPL Horizons program, based on an extrapolation of photometry obtained at small phase angles.  The approximate apparent visual magnitude is calculated for solid body reflection at the given heliocentric and Phaethon-STEREO distances. The first observations in each year are brighter than the prediction by about  0.5 magnitude.  This difference is physically insignificant, since the Horizons brightness prediction is based on an assumed (not measured) phase function extrapolated to large angles.  However, the brightening with time (and phase angle) shortly after perihelion in each year is highly significant.  Monolithic, macroscopic bodies  fade dramatically as a result of phase darkening in this range, opposite to the observed brightening.


\section{Discussion}

To correct for the variations in the heliocentric and Phaethon-STEREO distances, $R_{au}$ and $\Delta_{au}$, respectively (both expressed in AU), we use the inverse-square law, written

\begin{equation}
m(1,1,\alpha)=m_{obs}-5\log(R_{au}\Delta_{au}).
\label{m11a}
\end{equation}

\begin{figure}
\epsscale{1.0}
\begin{center}
\includegraphics[width=0.6\textwidth]{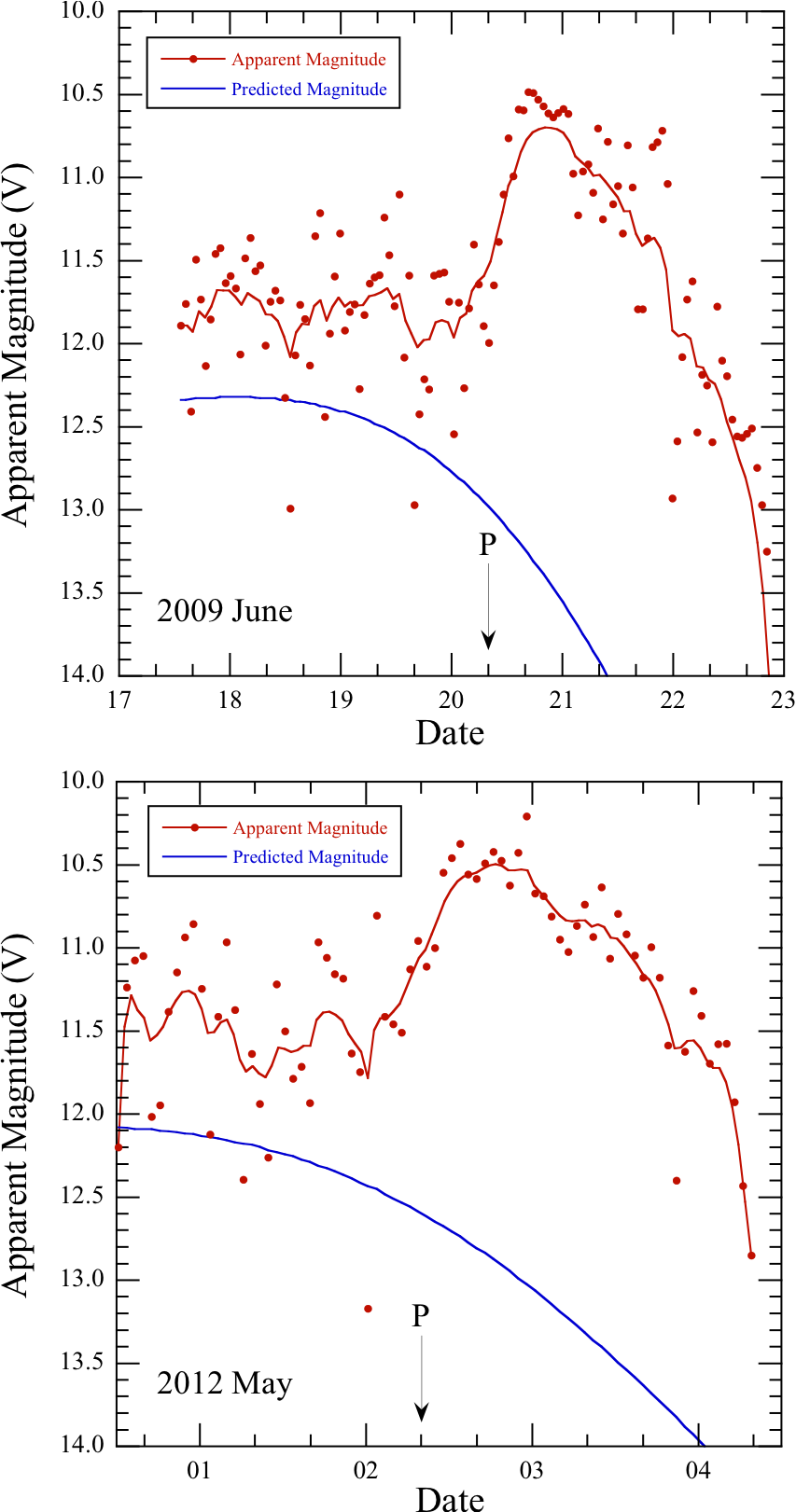}
\caption{Apparent $V$ magnitude of Phaethon (red circles) in 2009 and 2012. The magnitudes predicted by NASA's ephemeris software are shown for reference (blue lines). Letter ``P'' and arrows mark times of perihelion, date = 20.3 for June 2009 and date = 02.3 for May 2012. The red curve is a smoothed fit to the data added to guide the eye. The time ranges correspond to the valid Phaethon photometry measurements (see Table (\ref{stats})). \label{magv} }
\end{center} 
\end{figure}

\begin{figure}
\epsscale{0.9}
\begin{center}
\includegraphics[width=1.0\textwidth]{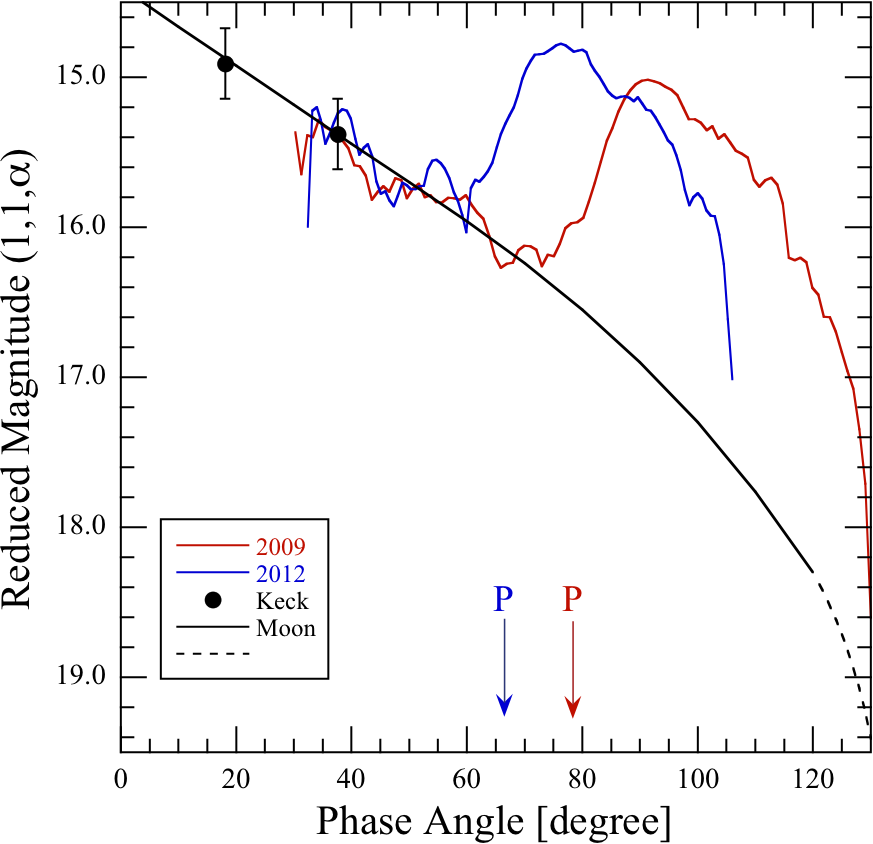}
\caption{Magnitude at $R$ = $\Delta$ = 1 AU, vs. phase angle ($\alpha$) in 2009 and 2012. The reduced Phaethon magnitudes are plotted in red (2009) and blue (2012) curves, and are smoothed fits to the actual data points. Two Keck data points are from a new measurement in 2012 for m(1,1,$18.1^\circ$); and from \citet{2005ApJ...624.1093H} for m(1,1,$37.6^\circ$). The lunar phase function is over-plotted from \citet{1973AJ.....78..267L} with the thick solid black curve.  For  $\alpha>120^\circ$, the lunar phase function is  extrapolated  (dashed curve). The letters ``P''  indicate phase angles $\alpha=79^\circ$ (red) and $\alpha=66^\circ$ (blue) corresponding to the perihelia in 2009 and 2012, respectively. Note that these perihelia correspond to the starts of the Phaethon brightnesses.  \label{mag-phase} }
\end{center} 
\end{figure}

\noindent Here $m_{obs}$ is the apparent magnitude and $m(1,1,\alpha)$ is the resulting magnitude which would be observed from $R_{au}$ = $\Delta_{au}$ = 1 and phase angle $\alpha$. Equation (\ref{m11a}) is plotted against $\alpha$ in Figure (\ref{mag-phase}).  Since the individual measurements are very scattered (Figure \ref{magv}), the  plotted curves show the running means of 10\% of the data.  Also plotted in the Figure is $m(1,1,18.1\degr)$ from our Keck observation on UT 2012 October 14, and $m(1,1,37.6\degr)$ from \citet{2005ApJ...624.1093H}.  The latter two measurements are plotted with error bars of $\pm$0.2 mag.~to represent uncertainty resulting from their being taken at unknown rotational phase.  

The combined ground-based and space-based data show a trend towards fading $m(1,1,\alpha)$ upto $\alpha \sim$ 60\degr.  Specifically, Phaethon fades by $\sim$1 mag. from $\alpha$ = 18.1\degr~to $\alpha$ = 60\degr, giving a linear phase coefficient, $\sim$0.024 mag.~degree$^{-1}$, that is unremarkable when compared with other asteroids.  At larger $\alpha$, Phaethon in Figure (\ref{mag-phase}) shows sudden distance-corrected brightening starting at $\alpha = 80^\circ$ in 2009 and $\alpha = 65\degr$ in 2012.  These phase angles correspond to perihelion in each year, while maximum brightness is reached at $\alpha = 100\degr$ in 2009 and $\alpha = 80\degr$ in 2012, about 0.5 day later. Phaethon fades to invisibility at larger phase angles.  As noted in Paper I, this brightness variation is unexpected for a solid body viewed in scattered light.  For such an object, the brightness decreases monotonically as the phase angle increases both because a progressively smaller fraction of the surface is illuminated and because the scattering efficiency of the surface decreases as the scattering angle grows.  

\begin{figure}
\epsscale{0.9}
\begin{center}
\includegraphics[width=1.0\textwidth]{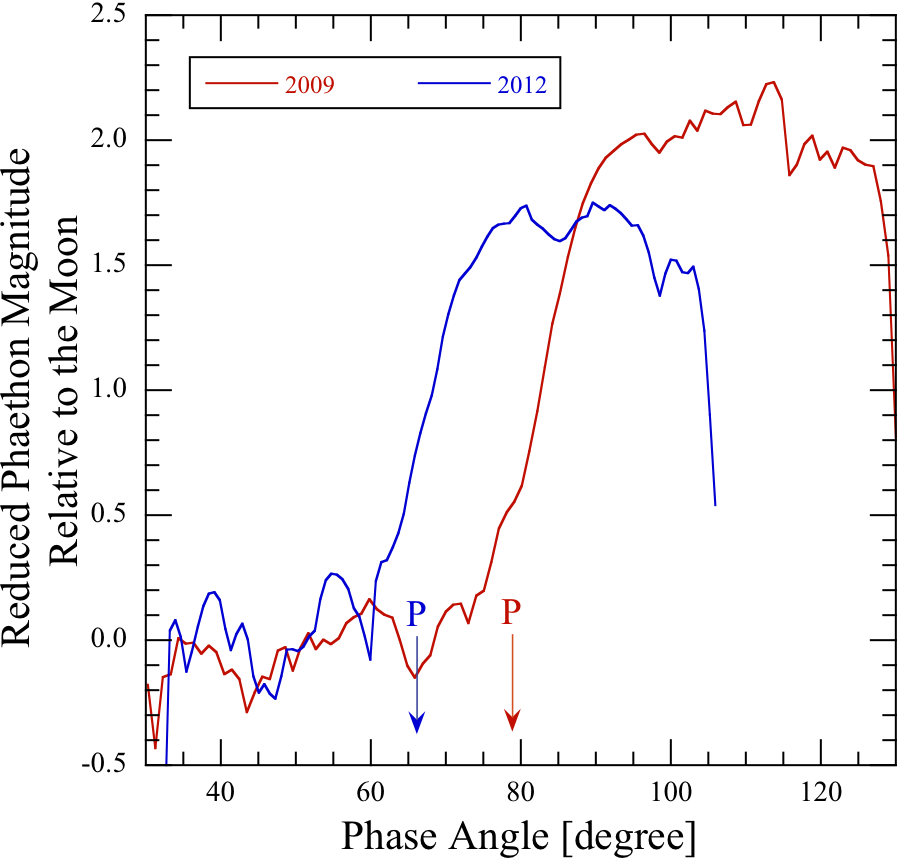}
\caption{Phase angle dependence normalized to the phase function of the Moon. The phase angles at perihelion are marked with arrows and the letter ``P''. \label{mag-phase_moon} }
\end{center} 
\end{figure}

This is emphasized in Figure (\ref{mag-phase}), which compares Phaethon with the scaled phase function of the Moon, taken from \citet{1973AJ.....78..267L}. The plotted lunar curve is an average of the phase functions at wavelengths 6264 \AA~and 7297 \AA~in Table V of their paper. We show the difference between $m(1,1,\alpha)$ and the scaled magnitude of the Moon in Figure (\ref{mag-phase_moon}).  The difference plot shows that Phaethon's phase dependence is Moon-like before the onset of the anomalous brightening event in each year.   While the apparent brightness of Phaethon (Figure \ref{magv}) increased by slightly more than a magnitude near $\alpha$ = 80\degr~to 100\degr, the brightening relative to the phase-darkened nucleus (Figure \ref{mag-phase}) is a much larger $\sim$2 magnitudes, corresponding to a factor $\sim$6.  The figure also shows that the apparent, post-peak fading in Figure (\ref{magv}) is consistent with the dimming expected from the phase function, not necessarily to loss of the scattering cross-section.

\subsection{Mechanisms}
We first consider and reject several mechanisms that might be implicated in the observed anomalous brightening at perihelion. 

The intrinsic brightening is too large (2 mag.)~and too long-lived (2 days) to be plausibly attributed to rotational variation of the projected cross-section of the aspherical nucleus of Phaethon.  Lightcurve observations show a variation $\le$0.4 magnitudes and a rotational period of only $\sim$3.6 hrs \citep{2002Icar..158..294K}.   The possibility that the brightening might be caused by a glint (a specular reflection from a mirror-like patch on the surface) can likewise be rejected on the basis of the longevity of the event. (Separately, the likelihood that the surface of a rocky asteroid could be mirror-like in the optical seems remote).   

As noted in Paper I, the brightening of Phaethon cannot be attributed to comet-like processes driven by the sublimation of near-surface water ice. This is because surface temperatures on Phaethon are far too high for water ice to survive.  The deep interior temperature (identified with the blackbody temperature of a body moving with Phaethon's orbitally-averaged heliocentric distance) is also too high to permit the survival of buried ice (Paper I).  Even if it were present, deeply-buried ice would be thermally decoupled from the instantaneous solar insolation (the conduction timescale across the radius is $\ge$10$^5$ yr), leaving no explanation for why the brightening is correlated with perihelion. 

The solar wind kinetic energy flux onto the surface of Phaethon is  $E_{KE} = \rho v^3/2$ (W m$^{-2}$), where $\rho$ is the solar wind mass density,  and $v$ is the solar wind speed. At the Earth's orbit, the solar wind number density is about 10$^7$ m$^{-3}$. Scaled by the inverse-square law to perihelion at 0.14 AU, the density is about $N_1$=5$\times$10$^8$  m$^{-3}$. The solar wind speed varies with time and radius, but is of order $v=500$ km/s. Substituting $\rho=\mu m_H N_1$, where $\mu=1$ (for protons), and $m_H=1.67\times 10^{-27}$ kg, we obtain $E_{KE} \sim$ 0.05 W m$^{-2}$. This is tiny compared with the solar radiation flux at perihelion, $F_{\odot}/R_{AU}^2=70,000$ W m$^{-2}$.  Consequently, we conclude that the solar wind is a negligible source of energy and cannot account for the anomalous brightening by impact fluorescence. 

Could part or all of the measured excess optical brightness be thermal emission resulting from Phaethon's high surface temperatures when at perihelion? We obtain a rough lower limit to the temperature by considering the case of an isothermal, spherical blackbody in equilibrium with sunlight, namely $T_{BB} = 278 R_{au}^{-1/2}$.  At perihelion, $R_{au}$ = 0.14, we find $T_{BB}$ = 743 K.  A practical upper limit is given by the sub-solar temperature on a non-rotating body (or a rotating one whose rotation axis points at the Sun), namely $T_{SS} = \sqrt{2} ~T_{BB}$, giving $T_{SS}$ = 1050 K for Phaethon at perihelion. This temperature range is in good agreement with independent estimates  ($\sim$800 $\le T \le$ 1100 K) from \citet{2009PASJ...61.1375O}.

A detailed calculation of the thermal emission from the surface of Phaethon depends on many unknowns and is beyond the scope of the present work.    Instead, we set a strong upper limit to the possible thermally emitted flux density by assuming that the whole surface of Phaethon (not just the sub-solar region) is at $T_{SS}$, the maximum possible surface equilibrium temperature. Under this assumption, the flux density in the STEREO bandpass is calculated from

\begin{equation}
f_\lambda = \frac{\pi r_n^2}{\Delta^2}\left[\frac{\int_{\lambda_1}^{\lambda_2}B_\lambda(T) d\lambda}{\Delta \lambda}\right]
\label{heat}
\end{equation}

\noindent in which $B_{\lambda}(T)$ is the Planck function evaluated at $T = T_{SS}$, $r_n$ = 2.5 km is the effective circular radius of Phaethon, $\Delta$ is the Phaethon-to-observer distance, the integration is taken over the filter transmission from $\lambda_1$ = 6300 \AA~to $\lambda_2$ = 7300 \AA, and $\Delta \lambda = \lambda_2 - \lambda_1$.   

\begin{figure}
\epsscale{1.0}
\begin{center}
\includegraphics[width=0.7\textwidth]{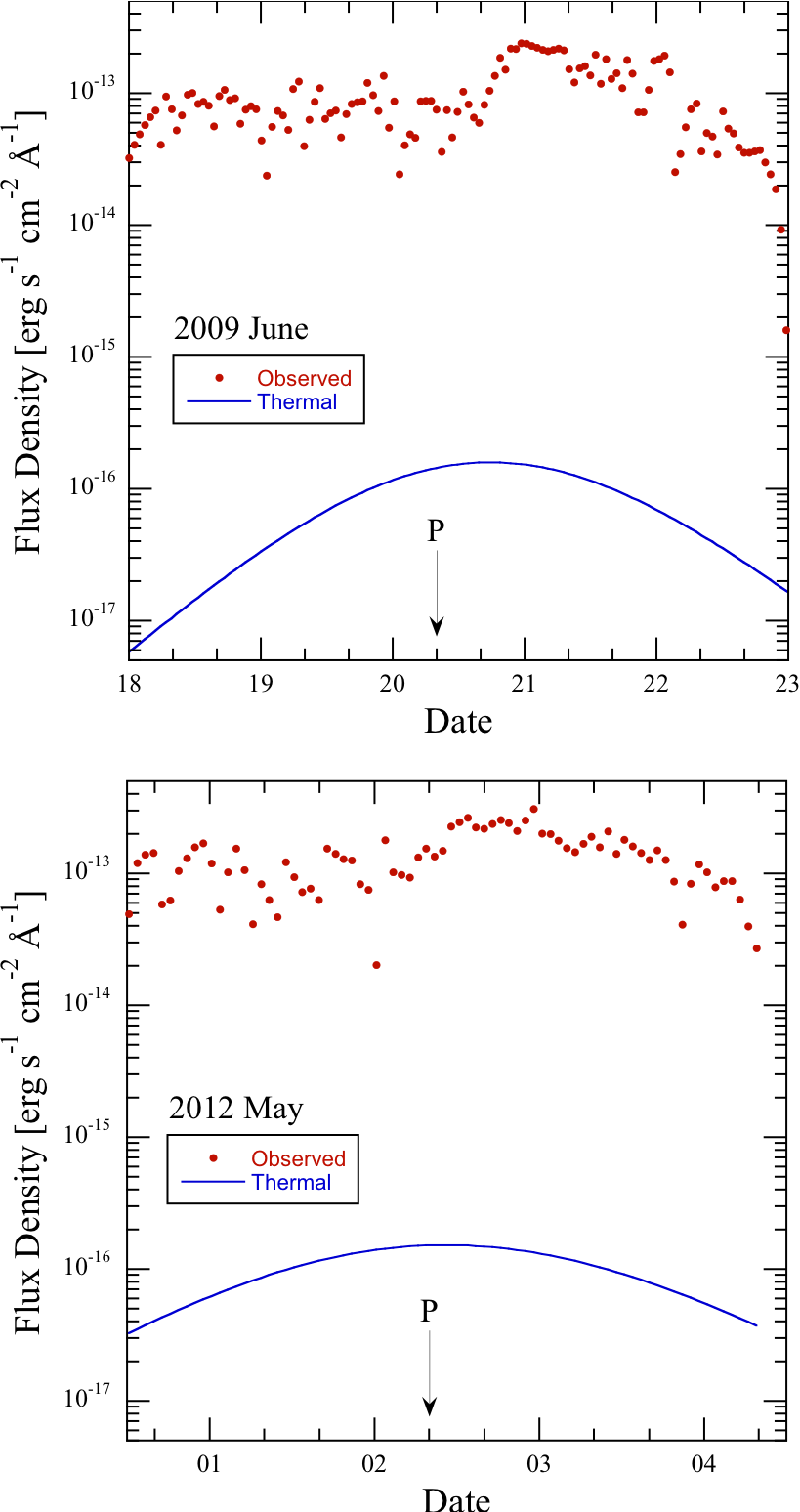}
\caption{Flux density as a function of time from photometry (red circles) and the thermal emission (blue curves). The latter is calculated from Equation (\ref{heat}) in 2009 (top) and 2012 (bottom).  The perihelia are marked by the letter ``P'' with arrows. \label{flux} }
\end{center} 
\end{figure}

For comparison with Equation (\ref{heat}), we convert the apparent magnitudes, $m_{obs}$, into flux densities, $f_\lambda^o$, using  $f_\lambda^o = 3.75\times 10^{-(9+m_{obs}/2.5)}$ [erg s$^{-1}$ cm$^{-2}$  \AA$^{-1}$] (Drilling and Landolt 2000). The results are shown in Figure (\ref{flux}), where solutions to Equation (\ref{heat}) (blue) are compared with the observations (red).  Evidently, thermal emission even at the peak sub-solar temperature is orders of magnitude too small to account for the anomalous brightening of Phaethon observed in our data.   

To see this a different way, we substituted the measured peak flux densities, $f_{\lambda}^o \sim$ 10$^{-13}$ erg s$^{-1}$ cm$^{-2}$ \AA$^{-1}$ (c.f. Figure \ref{flux}), into the left-hand side of Equation (\ref{heat}) and solved for the temperature. We find that values $T \sim$ 1650 K (2009) and 1700 K (2012) would be needed for thermal emission to account for the anomalous optical brightening.  This is far hotter even than the sub-solar temperature on the nucleus at perihelion and, therefore, can be discounted as unphysical.  We conclude that the perihelion brightening of Phaethon is not due to thermal emission from the surface.

The passband of the STEREO camera includes the 6300\AA~and 6363\AA~forbidden lines of oxygen.  These are ``prompt emissions'', formed when oxygen atoms are produced in the excited $^1$D state by photodissociation of a parent molecule, for example water (Festou and Feldman 1981).  As noted earlier, although water ice cannot survive, water might be bound within hydrated minerals in Phaethon and released by desiccation.  In the case of water, the photodissociation timescale at perihelion is about half an hour, so that any water molecules released from Phaethon would be destroyed within a single pixel of the STEREO camera.  We estimate the flux density produced by prompt emission in oxygen, averaged over the passband of the camera, from 

\begin{equation}
f_{\lambda}^{[OI]} = \frac{\alpha Q h c }{4\pi \Delta^2 \lambda \Delta \lambda }
\end{equation}

\noindent in which $\alpha \sim$ 10\% is the fraction of water dissociations leaving oxygen in the excited $^1$D state, $Q$ is the production rate of water molecules, $h$ is Planck's constant, $c$ is the speed of light, $\Delta$ is the Phaethon to STEREO distance, $\lambda$ is the wavelength and $\Delta \lambda$ is the filter FWHM, expressed in Angstroms.  Setting $f_{\lambda}^{[OI]}$ =  $f_{\lambda}^o$, we find that a water production rate $Q \sim$ 10$^{30}$ s$^{-1}$ (3$\times$10$^4$ kg s$^{-1}$) would be needed to account for the measured excess flux density of Phaethon.   Although the required rate of production (which is similar to that of comet 1P/Halley at perihelion) seems high, we cannot rigorously rule out the possibility that some fraction of the excess perihelion emission is caused by prompt emission from oxygen.   However, the observation that the fading of Phaethon after peak brightness follows the phase function of a solid object (Figure \ref{mag-phase}) suggests that gas is not the dominant cause of the anomalous brightness.

\subsection{Dust}
The remaining alternative is that Phaethon has ejected dust particles with a combined cross-section larger than that of the solid nucleus, as earlier concluded in Paper I.  In this scenario, the rise in brightness in Figure (\ref{photo}) then corresponds to the ejection of dust, while the subsequent decline in brightness is naturally explained as fading owing to the ever-growing phase dimming, perhaps aided by grain sublimation or disintegration.  The natural test of this hypothesis would be to search for coma scattered by the ejected dust.  Unfortunately, as also noted in Paper I, the limited angular resolution and high background surface brightness in the STEREO data make the detection of resolved coma impossible.

A temperature-controlled mechanism for the ejection of dust is strongly suggested;  the activity is observed at the highest (perihelion) temperatures, and is absent in observations of Phaethon taken at substantially larger heliocentric distances and lower temperatures.  The reflection spectrum of Phaethon has been interpreted in terms of thermally modified hydrated minerals \citep{2007A&A...461..751L,2009PASJ...61.1375O,2010A&A...513A..26D} while the depletion of sodium in some Geminids provides independent evidence for thermal alteration \citep{2006A&A...453L..17K}. The perihelion temperatures on Phaethon exceed those needed to break-down phyllosilicates \citep{1992AMR.....5..120A}, and are sufficient to induce thermal fracture \citep[Paper I,][]{2012AJ....143...66J}.   In this sense, thermal disintegration and fracture are plausible sources of the anomalous brightening and Phaethon may be accurately labeled a ``rock comet'' (Paper I).  An additional requirement is that dust must be cleared from the surface in order for these processes to operate.  A regolith of fine particles built up in previous orbits will inhibit thermal fracture, since small grains are unable to sustain large temperature differences.  Likewise, surface materials dehydrated by baking in previous perihelion passages must be cleared away from the surface in order for dehydration cracking to remain a persistent dust source.  

In both 2009 and 2012, the apparent brightness increased by $\sim$2 magnitudes relative to the nominal phase curve (Figures \ref{mag-phase} and \ref{mag-phase_moon}), corresponding to a factor of $\sim$6.  Given that the cross-section of the nucleus of Phaethon is $C_n$ = $\pi r_n^2 \sim$20 km$^2$, the cross-section of added dust is then $C$ = 100 km$^2$, in both years.  The mass in spherical particles of mean radius $\overline{a}$ having cross-section, $C$, is 

\begin{equation}
M_d \sim \frac{4 \rho \overline{a} C}{3} ,
\label{mass}
\end{equation}

\noindent where $\rho$ is the grain density.  With $\rho$ = 3000 kg m$^{-3}$, we find $M_d \sim$ 4$\times$10$^8$ $a_{mm}$ kg, where $a_{mm}$ is the grain radius expressed in millimeters. The mass of the nucleus, represented as a 2.5 km radius sphere of the same density, is a much larger 2$\times$10$^{14}$ kg.

The Geminid stream mass is 10$^{12} \le M_s \le$ 10$^{13}$ kg \citep{1989MNRAS.240...73H,1994A&A...287..990J}, while  the  Geminid stream lifetime estimated on dynamical grounds is $\tau \sim$1000 year  \citep{1989A&A...225..533G,2007MNRAS.375.1371R}.  Accordingly, to explain the entire mass in the Geminid stream through events like those observed here would require a number of similar events per orbit, $N$, given by

\begin{equation}
N \sim \left(\frac{M_s}{\tau}\right)  \left(\frac{P}{M_d}\right),
\end{equation}

\noindent where $P$ = 1.4 year is the orbital period.  Substituting, we obtain $N \sim$ 4$a_{mm}^{-1}$ to 40$a_{mm}^{-1}$.  If the particles are millimeter-sized, $a_{mm}$ = 1, then the Geminids could be supplied in steady-state by 4 $\le N \le$ 40 outbursts like the one observed, each orbit.    

However, there is no compelling physical reason to assume that Geminid stream production is in steady state.  While $a_{mm}$ = 1 may approximately represent the radii of the Geminid meteors, much larger examples up to 5 kg in mass (equivalent radius $\sim$7 cm) have been inferred from Lunar night-side impact flashes \citep{2008M&PSA..43.5169Y}.    Moreover, Phaethon is but part of the Phaethon-Geminid Complex which includes at least two other kilometer-scale bodies caused by fragmentation on an all-together much larger and longer (10$^6$ yr?) scale.  We conclude that, while continuing mass-loss near perihelion may contribute to actively replenishing the smaller Geminids, the PGC complex as a whole is likely the product of a more ancient and catastrophic breakup (c.f.~Jewitt and Hsieh 2006, Kasuga and Jewitt 2008).

\subsection{The Future}
Many puzzles remain in understanding the anomalous brightening of Phaethon and its relation to the ejection of dust and to the Geminids.  We list the following key questions:

\begin{enumerate}
\item What is the value of the effective dust radius, $a_{mm}$?  This radius directly affects estimates of the ejected mass through Equation (\ref{mass}) and so determines the extent to which on-going activity in Phaethon contributes to the Geminid stream. 


\item What is the origin of the $\sim$0.5 day phase lag between perihelion and the anomalous brightening in both 2009 and 2012?   Is the brightening always lagged relative to perihelion by this amount?  

\item How many brightening events occur per orbit and are these events always of the same amplitude?  The nature of the  STEREO data curtails our ability to detect brightening events outside a limited window of accessibility.

\item Can evidence for mass loss be detected when Phaethon is far from perihelion, despite past, failed attempts? Thermal fracture and mineral decomposition are likely only at the extreme temperatures found near perihelion.  However, slow-moving dust may linger in the post-perihelion months, making these the prime time for future attempted dust observations.

\item Can spectra be obtained at perihelion in order to search for the forbidden lines of oxygen?

\item What is the mineralogical composition of Phaethon and are hydrated silicates present on its surface?  

\item Can meteoroids ejected during recent perihelion events be detected and distinguished from older Geminids? \citet{2012MNRAS.423.2254R} reports that this will be difficult.

\end{enumerate}

\clearpage

\section{Summary}

We report new observations of planet-crossing asteroid and Geminid meteoroid parent (3200) Phaethon, using the NASA STEREO solar spacecraft.  We find that

\begin{enumerate}

\item  (3200) Phaethon exhibited anomalous brightening when at perihelion in 2009 and 2012, but not in 2010 (the latter likely owing to unfavorable observing geometry). The distance-corrected apparent brightness increased near phase angle 100\degr~in 2009 and 80\degr~in 2012,  in both years $\sim$0.5 day after perihelion passage.  This brightening lies in stark contrast to the monotonic fading expected from phase darkening on a macroscopic body. 

\item  The most direct interpretation is that Phaethon brightens because of a sudden increase in the scattering cross-section due to the ejection of dust with a mass  $M_d \sim$ 4$\times$10$^8 a_{mm}$ kg, where $a_{mm}$ is the effective dust radius in millimeters.  A contribution from prompt emission by atomic oxygen cannot be excluded.

\item Thermal fracture and the decomposition of hydrated silicates are two plausible mechanisms of dust production at the $\sim$1000 K surface temperatures attained near perihelion. Both are difficult to quantify in the absence of more detailed information about the composition of Phaethon.

\item Phaethon has only very limited visibility in the STEREO field of view (typically $\le$5 days per orbit).  The detection of anomalous brightening twice in two favorable observing windows suggests that this phenomenon is common.

\end{enumerate}

\acknowledgments
We thank Man-To Hui for alerting us to the 2012 appearance of Phaethon. Bill Thompson at NASA promptly and kindly answered our inquiry on the celestial position conversions to the image pixels.  Bin Yang, Toshi Kasuga and the anonymous referee offered helpful comments.  We thank Mike A'Hearn for suggesting that we consider forbidden oxygen. The Heliospheric Imager instrument was developed by a collaboration that included the University of Birmingham and the Rutherford Appleton Laboratory, both in the UK, the Centre Spatial de Liege (CSL), Belgium, and the U.S. Naval Research Laboratory (NRL), Washington DC, USA. The STEREO/SECCHI project is an international collaboration. Some of the data presented herein were obtained at the W.M. Keck Observatory, which is operated as a scientific partnership among the California Institute of Technology, the University of California and the National Aeronautics and Space Administration. The Observatory was made possible by the generous financial support of the W.M. Keck Foundation.  This work was supported in part by a grant from the NASA Planetary Astronomy Program.

\clearpage

\begin{deluxetable}{clccccc}
\tablecaption{Photometry Statistics and Detection Characteristics
\label{stats}}
\tablewidth{0pt}
\tablehead{
\colhead{Year} &
\colhead{Target} &
\colhead{Mean} &
\colhead{Std. Deviation} &
\colhead{Threshold\tablenotemark{\dag}} &
\colhead{Date\tablenotemark{\ddag}} &
\colhead{Phase Angle\tablenotemark{\S}} }
\startdata
2009 & Sky & 0.09 & 0.04& \nodata & \nodata & \nodata \\
 & Phaethon & 0.44 & 0.37 & 0.27 &June 17.5-23.0 & $29^\circ$-$130^\circ$ \\
 \tableline
 2012 & Sky & 0.10 & 0.12 &\nodata & \nodata & \nodata \\
 & Phaethon & 0.36 & 0.40 & 0.30  & April 30.5 - May 4.3 & $32^\circ$-$105^\circ$\\
\enddata
\tablenotetext{\dag}{The adopted threshold for Phaethon photometry, equal to three times the sky mean. }
\tablenotetext{\ddag}{Time interval during which Phaethon's light curves are above the threshold.}
\tablenotetext{\S}{Phase angle ranges corresponding to these date ranges.}
\end{deluxetable}

\begin{deluxetable}{lclllcl}
\tablecaption{Standard Stars
\label{stars}}
\tablewidth{0pt}
\tablehead{
\colhead{} &
\colhead{HIC} &
\colhead{HD} &
\colhead{SAO} &
\colhead{[R.A., Dec.]} &
\colhead{Magnitude} &
\colhead{Spectra} }
\startdata
\multicolumn{7}{l}{\bf 2009}\\
       1&       47142&       83099&       98670&       144.0,       10.6&      8.50&F2         \\
       2&       48246&       85143&      117932&       147.5,       9.7&      8.48&G5         \\
       3&       48381&           \nodata&       98815&       147.9,       9.8&      8.40&K0         \\
       4&       48624&          \nodata &       98843&       148.7,       9.9&      8.40&F8         \\
       5&       48866&       86340&      118022&       149.5,       9.7&      8.50&F5         \\
       6&       49194&        \nodata   &       98915&       150.6,       9.9&      8.97&K0         \\
       7&       50146&       88724&       99018&       153.6,       10.5&      9.40&G          \\
       8&       50110&       88680&       99012&       153.5,       10.8&      8.02&G5         \\
       9&       50424&       89209&       99041&       154.4,       11.3&      8.60&F5         \\
      10&       50467&         \nodata  &       \nodata    &       154.6       11.8&      9.30&K0         \\
      11&       50911&       90042&       99096&       155.9,       14.8&      8.58&G5         \\
\tableline
\multicolumn{7}{l}{\bf 2012}\\
       1&       52259&       92458&      118427&       160.2,       5.5&      8.90&G5         \\
       2&       53032&       93981&      118531&       162.8,       5.0&      8.68&K5         \\
       3&       53165&       94220&      118554&       163.1,       5.0&      8.17&K0         \\
       4&       53899&       95529&      118622&       165.4,       5.1&      8.80&G5         \\
       5&       54120&           \nodata&      118647&       166.1,       5.1&      9.10&F8         \\
       6&       54213&           \nodata&      118652&       166.4,       4.9&      8.70&K0         \\
       7&       54188&           \nodata&      118649&       166.3,       5.2&      9.61&F8V        \\
       8&       54481&       96762&      118684&       167.2,       5.0&      8.83&F0         \\
       9&       54660&           \nodata&      118707&       167.8,       5.1&      9.30&F8         \\
      10&       54788&       97459&      118724&       168.2,       5.2&      8.50&G0         \\
      11&       55323&       98451&      118791&       169.9,       6.2&      9.20&K0         \\
      12&       56338&      100362&      118913&       173.2,       9.4&      8.38&F8         \\
      13&       56542&      100727&      118935&       173.9,       9.6&      8.35&F2         \\
      14&       56491&           \nodata&           \nodata&       173.7,       10.0&      9.70&F2         \\
      15&       56646&      100904&       99697&       174.2,      10.5&      8.26&F5         \\
\enddata
\end{deluxetable}

\clearpage

\clearpage

\clearpage

\clearpage

\clearpage

\clearpage

\clearpage
%

\end{document}